\documentclass[twocolumn,showpacs,preprintnumbers,nofootinbib,amsmath,amssymb]{revtex4}

\usepackage{euscript,amssymb,amsmath}

\usepackage{graphicx}
\usepackage{epsfig}

\newcommand{\be}[1]{\begin{equation}\label{#1}}
\newcommand{\ee}{\end{equation}}
\newcommand{\ba}[1]{\begin{eqnarray}\label{#1}}
\newcommand{\ea}{\end{eqnarray}}
\newcommand{\rf}[1]{(\ref{#1})}
\newcommand{\nn}{\nonumber}

\begin{document}

%%%%%
\title{The shape of multidimensional gravity}
%%%%

\author{Maxim Eingorn}\email{maxim@scientist.com}  \author{Alexander Zhuk}\email{ai_zhuk2@rambler.ru}
\affiliation{Astronomical Observatory and Department of
Theoretical Physics, Odessa National University, Street Dvoryanskaya 2, Odessa 65082, Ukraine}

%\date{23 March 2007}
%
%%%%%%%%%%%%%%%%%%%%%%%%%%%%%%%%%%%%%%%%%%%%%%%%%%%%%%%%%
%
%

%
\begin{abstract} In the case of one extra dimension, well known Newton's potential $\varphi (r_3)=-G_N m/r_3$ is generalized to compact and elegant
formula $\varphi(r_3,\xi)=-(G_N m/r_3)\sinh(2\pi r_3/a)[\cosh(2\pi r_3/a)-\cos(2\pi\xi/a)]^{-1}$ if four-dimensional
space has topology $\mathbb{R}^3\times T$. Here, $r_3$ is magnitude of three-dimensional radius vector, $\xi$ is extra dimension and $a$ is a period of a torus $T$. This formula is valid for full range of variables $r_3 \in [0,+\infty )$ and $\xi\in [0,a]$ and has known asymptotic behavior: $\varphi \sim 1/r_3$ for $r_3>>a$ and $\varphi
\sim 1/r_4^2$ for $r_4=\sqrt{r_3^2+\xi^2}<<a$. Obtained formula is applied to an infinitesimally thin shell, a shell, a sphere and two spheres to show deviations from the newtonian expressions. Usually, these corrections are very small to observe at experiments. Nevertheless, in the case of spatial topology
$\mathbb{R}^3\times T^{d}$,
experimental data can provide us with a limitation on maximal number of extra dimensions.
\end{abstract}

\maketitle

\vspace{.5cm}

%\today \\
%\mbox{}\\
%%%%%%%%%%%%%%%%%%%%%%%%%%%%%%%%%%%%%%%%%%%%%%%%%%%%%%%%%%%%%
%\section{Introduction}
%\setcounter{equation}{0}

%\vspace{1cm}

{\em Introduction}\quad {}From school years we know that newtonian gravitational potential of a body with mass $m$ has the form of
$\varphi(r) = -G_N m /r$
where $G_N$ is Newton's gravitational constant
and $r = |{\bf r}|$ is magnitude of a
radius vector in three-dimensional space.
This expression can be derived from Gauss's flux theorem in three-dimensional space.
However, it is possible that spacetime might have a dimensionality of more than four and posses a rather complex topology.
String theory and its supersymmetric generalizations - superstring and supergravity widely use this concept and give it
a new foundation. Therefore, it is of interest to generalize the Newton's
formula to the case of extra dimensions. Clearly, the result depends on topology of investigated models.
We consider models where $(D=3+d)$-dimensional spatial part of factorizable geometry is defined on a product manifold
$M_D=\mathbb{R}^3\times T^{d}$.
$\mathbb{R}^3$ describes three-dimensional flat external (our) space and $T^{d}$ is a torus
which corresponds to $d$-dimensional internal space. Let $V_{d}$ be a volume of the internal space and $b\sim V^{1/d}_{d}$
is a characteristic size of extra dimensions. Then, Gauss's flux theorem leads to the following asymptotes for
gravitational potential (see e.g. \cite{ADD}): $\varphi \sim 1/r_3$ for $r_3 >> b$ and $\varphi \sim 1/r^{1+d}_{3+d}$
for  $r_{3+d} << b$ where $r_3$ and $r_{3+d}$ are magnitudes of radius vectors in three-dimensional and
$(3+d)$-dimensional spaces, respectively.
%We shall consider $D=4+d$-dimensional factorizable geometry with a product manifold (we show here only spatial part)
%$M_D=\mathbb{R}\times\mathbb{R}^3\times T^{d}$ where
%$\mathbb{R}\times\mathbb{R}^3$ describes four-dimensional flat external (our) spacetime and $T^{d}$ is a torus
%which describes $d$-dimensional internal space. Let $V_{d}$ be a volume of the internal space and $b\sim V^{1/d}_{d}$
%is a characteristic size of extra dimensions. Then, Gauss's flux theorem results in the following asymptotes for
%gravitational potential (see e.g. \cite{ADD}):
%%%%%%%
%\be{1}
%\varphi
%%_{3}(r_3)
%\propto -G_N m/r_3\, , \quad  r_3 >> b
%\ee
%%%%%%%
%and
%%%%%%%
%\be{2}
%\varphi
%%_{3+d}(r_{d+3})
%\propto -G_D m / r^{1+d}_{3+d}\, , \quad  r_3 << b\, ,
%\ee
%%%%%%
%where $G_D$ is gravitational constant in $D$-dimensional spacetime and $r_3$ and $r_{d+3}$ are radius vectors in three-dimensional and
%$(d+3)$-dimensional spaces, respectively.
%The main purpose of our paper consists in generalization of the Newton's expression to multidimensional case in the form of
%a compact elegant formula with the shown above asymptotes.
In our letter, we generalize the Newton's expression to multidimensional case with the shown above asymptotes
and demonstrate that at least in the case $d=1$ it has the form of compact and elegant formula via elementary functions.
%search of a compact elegant formula for gravitational potential which generalizes the
%Newton's expression to multidimensional case and which has the shown above asymptotes. In this letter,
%we demonstrate that such possibility exists at least in the case $d=1$.
%The exact expression has the form of infinite series over all Kaluza-Klein modes \cite{ADD,CB}.
%Is it possible to sum up these series to get compact and elegant formula via elementary functions? In this letter,
%we shall show that such possibility exists at least in the case $d=1$.

%%%%%%%%%%%%%%%%%%%%%%%%%%%%%%%%%%%%%%%%%%%%%%%%%%%%%%%%%%%%%%%%%%%%%%%%%%%%

%\section{Linear model}
%\setcounter{equation}{0}

%\vspace{1cm}

{\em Multidimensional gravitational potentials}\quad In $D$-dimensional space,
the  Poisson equation reads
%%%%
\be{1}
\triangle_D\varphi_D=S_DG_{\mathcal{D}}\rho_D({\bf r}_D)\, ,
\ee
%%%%%
where $S_D=2\pi^{D/2}/\Gamma (D/2)$ is a total solid angle (square of
$(D-1)$-dimensional sphere of a unit radius), $G_{\mathcal{D}}$ is a gravitational constant in
$(\mathcal{D}=D+1)$-dimensional spacetime
and $\rho_D({\bf r}_D)=m\delta(x_1)\delta(x_2)...\delta(x_D)$. In the case of topology
$\mathbb{R}^D$, Eq. \rf{1} has the following solution:
%%%%%%%
\be{2}
\varphi_D({\bf r}_D)=-\frac{G_{\mathcal{D}}m}{(D-2)r_D^{D-2}}\, ,\quad D\geq 3.
\ee
%%%%%%%
This is the unique solution of Eq. \rf{1} which satisfies the boundary condition:
$\lim\limits_{r_D\rightarrow+\infty}\varphi_D({\bf r}_D)=0$. Gravitational constant
$G_{\mathcal{D}}$ in \rf{1} is normalized in such a way that the strength of gravitational
field (acceleration of a test body) takes the form: $-d\varphi_D / d r_D = - G_{\mathcal{D}}m/r^{D-1}_D$.

If topology of space is  $\mathbb{R}^3\times T^{d}$, then it is natural to impose periodic boundary
conditions in the directions of the extra dimensions:
$\varphi_D({\bf r}_3,\xi_1,\xi_2,\ldots, \xi_i,\ldots ,\xi_{d})=
\varphi_D({\bf r}_3,\xi_1,\xi_2,\ldots, \xi_i +a_i,\ldots ,\xi_{d}), \quad i=1,\ldots ,d$, where
$a_i$ denotes a period in the direction of the extra dimension $\xi_i$. Then, Poisson equation has solution
(cf. also with \cite{ADD,CB}):
%%%%%%
\ba{3}
&{}&\varphi_D({\bf r}_3,\xi_1,...,\xi_{d})=-\frac{G_N m}{r_3}\nn \\
&\times&\sum\limits_{k_1=-\infty}^{+\infty}...\sum\limits_{k_{d}=-\infty}^{+\infty}
\exp\left[-2\pi\left(\sum\limits_{i=1}^{d}\left(\frac{k_i}{a_i}\right)^2\right)^{1/2}r_3\right]\nn \\
&\times&\cos\left(\frac{2\pi
k_1}{a_1}\xi_1\right)...\cos\left(\frac{2\pi k_{d}}{a_{d}}\xi_{d}\right)\, .
\ea
%%%%%%%
To get this result we, first, use the formula
$\delta(\xi_i)=\frac{1}{a_i}\sum_{k=-\infty}^{+\infty}\cos\left(\frac{2\pi k}{a_i}\xi_i\right)$ and, second,
put the following relation between gravitational constants in four- and $\mathcal{D}$-dimensional spacetimes:
%%%%%
\be{4}
\frac{S_D}{S_3}\cdot\frac{G_{\mathcal{D}}}{\prod_{i=1}^{d}a_i}=G_N\, .
\ee
%%%%%
The letter relation provides correct limit when all $a_i \to 0$. In this limit zero modes $k_i=0$ give the main
contribution and we obtain $\varphi_D({\bf r}_3,\xi_1,...,\xi_{d})\rightarrow-G_N m/r_3$. Eq. \rf{4} was
widely used in the concept of large extra dimensions which gives possibility to solve the hierarchy problem \cite{ADD,large}.
It is also convenient to rewrite \rf{4} via fundamental Planck scales:
%%%%%%%
\be{5}
\frac{S_D}{S_3}\cdot M_{Pl(4)}^{2} = M_{Pl(\mathcal{D})}^{2+d}\prod_{i=1}^{d}a_i\, ,
\ee
%%%%%%%
where $M_{Pl(4)}= G_N^{-1/2} =1.2\cdot 10^{19}$GeV and $ M_{Pl(\mathcal{D})}\equiv G_{\mathcal{D}}^{-1/(2+d)}$ are
fundamental Planck scales in four and $\mathcal{D}$ spacetime dimensions, respectively.

In opposite limit when all $a_i \to +\infty$ the sums in Eq. \rf{3} can be replaced by integrals. Using the standard integrals
(e.g. from \cite{PBM}) and relation \rf{4}, we can easily show  that, for example, in particular cases $d=1,2$ we get
desire result: $\varphi_D({\bf r}_3,\xi_1,\ldots ,\xi_d)\rightarrow-G_{\mathcal{D}} m/[(D-2)\; r_{3+d}^{1+d}]$.

{\em One extra dimension}\quad In the case of one extra dimension $d=1$ we can perform summation of series in Eq. \rf{3}.
To do it, we can apply the Abel-Plana formula or simply use the tables of series \cite{PBM}. As a result, we arrive at
compact and nice expression:
%%%%%
\be{6}
\varphi_4({\bf r}_3,\xi)=-\frac{G_N m}{r_3}\frac{\sinh\left(\frac{2\pi r_3}{a}\right)}{\cosh\left(\frac{2\pi
r_3}{a}\right)-\cos\left(\frac{2\pi\xi}{a}\right)}\, ,
\ee
%%%%%
where $r_3 \in [0,+\infty )$ and $\xi \in [0,a]$. It is not difficult to verify that this formula has correct asymptotes
when $r_3>>a$ and $r_4<<a$. Fig. 1 demonstrates the shape of this potential. Dimensionless variables $\eta_1\equiv r_3/a \in [0,+\infty )$
and $\eta_2\equiv \xi/a \in [0,1]$. With respect to variable $\eta_2$, this potential has two minima at $\eta_2=0,1$ and one maximum at
$\eta_2 =1/2$. We continue the graph to negative values of $\eta_2 \in [-1,1]$ to show in more detail
the form of minimum at $\eta_2=0$. The potential \rf{6} is finite for any value of $r_3$ if $\xi\neq 0,a$ and goes
to $-\infty $ as $ -1/r_4^2$ if
simultaneously $r_3 \to 0$ and $\xi \to 0,a$ (see Fig. 2).
%%%%%%%%%%%%%%%%%%%%%
\begin{figure}[htbp]
%\centerline{
\includegraphics[width=2.7in,height=1.8in]{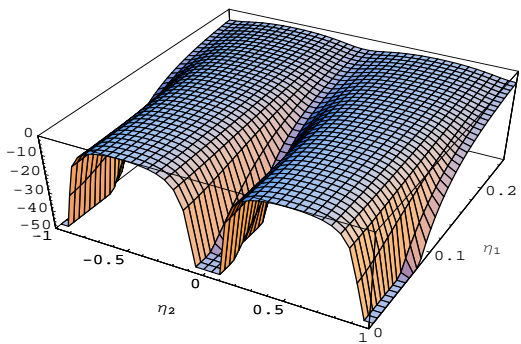}
\caption {Graph of function $\tilde \varphi (\eta_1,\eta_2) \equiv \varphi_4({\bf r}_3,\xi)/(G_N m/a)=
-\sinh (2\pi \eta_1)/[\eta_1(\cosh(2\pi\eta_1)-\cos(2\pi\eta_2))]$. \label{potential}}
\end{figure}
%%%%%%%%%%%%%%%%%%%%%%%%%%%%

%%%%%%%%%%%%%%%%%%%%%
\begin{figure}[htbp]
%\centerline{
\includegraphics[width=2.5in,height=1.4in]{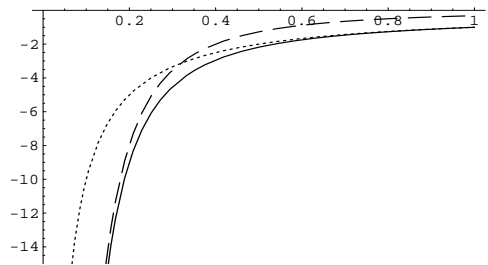}
\caption {Section $\xi =0$ of potential \rf{6}. Solid line is $\tilde \varphi (\eta_1,0) =
-\sinh (2\pi \eta_1)/[\eta_1(\cosh(2\pi\eta_1)-1)]$ which goes to $-1/\eta_1$ (dotted line) for $\eta_1 \rightarrow +\infty$
and to $-1/(\pi \eta_1^2)$ (dashed line) for $\eta_1 \rightarrow 0$. \label{asympt}}
\end{figure}
%%%%%%%%%%%%%%%%%%%%%%%%%%%%

Having at hand formula \rf{6}, we can apply it for calculation of some elementary physical problems and compare
obtained results with famous newtonian expressions.
For our calculations we shall use the case of $\xi =0$. It means that test bodies have the same coordinates in extra dimensions.
It takes place e.g. when test bodies are on the same brane. Also, to get  numerical results we should define the size $a$ of the
extra dimension. If the standard model fields are not localized on the brane, then experiments
give upper bound $a\lesssim 10^{-17}$cm \cite{inverse-square law}. For this value of $a$ the 5-dimensional fundamental Planck scale is
$M_{Pl(5)}\gtrsim 10^{11}$TeV (see Eq. \rf{5}). This value $a$ can be greatly increased if we suppose that the standard model
fields are localized on the brane. The gravitational inverse-square law experiments show that
there is no deviations from three-dimensional  Newton's law up to $2.18 \cdot 10^{-2}$cm \cite{Hoyle}. Thus, we can take for this model $a \approx 10^{-2}\div 10^{-3}$cm
which results in $M_{Pl(5)} \approx 10^6$TeV. However, it is necessary to keep in mind that this case can be
%strongly
constrained by observations from supernova cooling (see e.g. \cite{supernova}).

It is worth of noting that all formulas in this letter are applied to the Coulomb's law if electromagnetic field is not
localized on the brane. It was not found deviations from three-dimensional Coulomb's law up to
$10^{-16}$cm \cite{inverse-square law}.
%From experiments, it is known that in three-dimensional space there is no deviation from the Coulomb's law up to
%$10^{-16}$cm \cite{inverse-square law}.
Therefore, for models with non-localized electromagnetic field we should take $a\lesssim 10^{-17}$cm.

{\em Infinitesimally thin shell} \quad Let us consider first an infinitesimally thin shell of mass $m=4\pi R^2 \sigma$, where $R$ and $\sigma$ are  radius and surface density of
the shell. Then, gravitational potential of this shell in a point with radius vector $\bf r_3$ (from the center of the shell) is
%%%%%%
\be{7}
\varphi(r_3)=-\frac{G_N\sigma
Ra}{r_3}\ln\left\{\frac{\cosh\left(\frac{2\pi(r_3+R)}{a}\right)-1}{\cosh\left(\frac{2\pi(r_3-R)}{a}\right)-1}\right\}\, .
\ee
%%%%%%
%where $\varphi(r_3)\equiv \varphi_4({\bf{r_3}},0)$.
This formula demonstrates two features of the considered models. Firstly, we see that inside $(r_3<R)$ of the shell  gravitational
potential is not a constant. Thus, a test body undergoes an acceleration in contrast to the newtonian case (and Birkhoff's theorem
of general relativity in four-dimensional spacetime which states that the metric inside an empty spherical cavity in the center of a spherically symmetric system is the Minkowski metric). Secondly, this potential has a logarithmic divergency when $r_3 \to R$:
$\varphi(r_3)\approx -\frac{G_N m}{R}\left[1-\frac{a}{2\pi
R}\ln\left(2\pi|R-r_3|/a\right)\right]$  where we took into account $R>>a$ and $|R-r_3|<<a$. For example, in the case $2\pi R =10$cm and
$2\pi |R-r_3| = 10^{-1}a$, the deviation constitutes $2.3\cdot 10^{-4}$ and $2.3\cdot 10^{-18}$ parts of the newtonian value $-G_Nm/R$ for $a = 10^{-3}$cm and $a=10^{-17}$cm, respectively.
In principle, the former estimate is not very small. However, it is very difficult to set an experiment which satisfies the condition
$|R-r_3|<<a$.
%It is hardly possible to observe such deviations in experiments.
If the shell has a finite thickness, then the divergence disappears.

{\em Spherical shell}\quad The gravitational potential of a spherical shell of inner radius $R_1$ and outer radius $R_2$ can be written as
%%%%
\be{8}
\varphi(r_3)=-\frac{G_N\rho a}{r_3}\int\limits_{R_1}^{R_2}
R\ln\left\{\frac{\cosh\left(\frac{2\pi(r_3+R)}{a}\right)-1}{\cosh\left(\frac{2\pi(r_3-R)}{a}\right)-1}\right\}dR\, ,
\ee
%%%%%
where $\rho$ is a constant volume density of the shell: $\rho = m /\left[\frac{4\pi}{3}(R_2^3-R_1^3)\right]$.
It is useful to present this potential in the form of series. For example, inside $(r_3\leq R_1)$ of the shell it reads:
%%%%%%
\ba{9}
\varphi(r_3)&=&2G_N\rho\left\{-\pi \left(R_2^2-R_1^2\right)
+ \frac{a^2}{\pi r_3}\sum_{k=1}^{+\infty}\frac{\sinh\left(\frac{2\pi k}{a}r_3\right)}{k^2}\right.\nn\\
&{\times}&\left.\left[\left(R
+\frac{a}{2\pi k}\right)\exp\left(-\frac{2\pi k}{a}R\right)\right]_{R_1}^{R_2}\right\}\, ,
\ea
%%%%%%%
where we singled out zero mode $k=0$ which corresponds to the newtonian limit. It can be easily seen that this
series does not diverge when $r_3 \to R_1$. However, acceleration of a test body diverges when it approaches boundaries
$R_1$ and $R_2$ of the shell. It clearly follows e.g. from the form of the acceleration inside $(r_3\leq R_1)$ of the shell:
%%%%%
\ba{10}
&-&\frac{d\varphi}{dr_3}=\frac{2G_N\rho a^2}{\pi
r_3^2}\sum_{k=1}^{+\infty}\frac{1}{k^2}\left[-\frac{2\pi k}{a}r_3\cosh\left(
\frac{2\pi k}{a}r_3\right)\right. \nn \\
&+& \left.\sinh\left(\frac{2\pi k}{a}r_3\right)\right]
\left[\left(R+\frac{a}{2\pi k}\right)\exp\left(-\frac{2\pi k}{a}R\right)\right]_{R_1}^{R_2}\geqslant0\nn\\
&{}&
\ea
%%%%%
and outside $(r_3\geq R_2)$ of the shell:
%%%%%
\ba{11}
&-&\frac{d\varphi_n}{dr_3}=-\frac{G_N
m}{r_3^2}-\frac{G_N\rho a^3}{\pi^2r_3^2}\sum_{k=1}^{+\infty}\frac{
\left(1+\frac{2\pi k}{a}r_3\right)}{k^3}\nn \\
&\times& \exp\left(-\frac{2\pi k}{a}r_3\right)\left[h_k(R_2)-h_k(R_1)\right]<0\, ,
\ea
%%%%%%
where $h_k(R)=\frac{2\pi k}{a}R\cosh(\frac{2\pi k}{a}R)-\sinh(\frac{2\pi k}{a}R)$.
%%%%
Divergence of acceleration originates from the divergence of the series $\sum_{k=1}^{+\infty}1/k$ and has the form
$\pm 2G_N\rho a \ln\frac{2\pi\varepsilon}{a}$. Here, $\varepsilon = |R_{1,2}-r_3|$ and $-,+$ corresponds to $r_3\to R_1$
and $r_3\to R_2$,
respectively. In the case of a sphere ($R_1=0$ and $R_2\equiv R$) this divergence can be rewritten in the form
$-3\frac{G_Nm}{R^2}\frac{a}{2\pi R}\ln(2\pi|R-r_3|/a)$. Similar to the case of the infinitesimally thin shell, this deviation from the
newtonian acceleration
$-G_Nm/R^2$ is also difficult to observe
at experiments for considered above parameters. Eqs. \rf{10} and \rf{11} show that acceleration changes the sign from negative
outside of the shell to positive inside of the shell. This change happens within the shell. In the limit $R_1\to 0$,
some of obtained above formulas can be applied for a solid sphere.

{\em Gravitational self-energy of a sphere} \quad For a sphere of constant volume density $\rho$, mass $m$ and radius $R$, a gravitational
self-energy is
%%%%
\ba{12}
U&=&
%-\frac{3G_N m^2}{5R}
U_N\left\{1 \phantom{\int}\right.\\
&+&\left.15 \left(\frac{a}{2\pi R}\right)^3\sum\limits_{k=1}^{+\infty}
\frac{1}{k^3}\left[\frac{2\pi k R}{3a}+\left(\frac{a}{2\pi k R}\right)^2F_k\right]\right\}\nn \, ,
\ea
%%%%%
where
%%%%
\ba{13}
F_k&=&\left(1+\frac{2\pi k R}{a}\right) \exp\left(-\frac{2\pi k R}{a}\right)
\left[\sinh\left(\frac{2\pi k R}{a}\right)\right.\nn \\
&-&\left.\frac{2\pi k R}{a}\cosh\left(\frac{2\pi k R}{a}\right)\right]\nn
\ea
%%%%
and $U_N=-3G_N m^2/(5R)$
is gravitational self-energy of a sphere in the newtonian limit \cite{Kittel}.
In the case $R>>a$, the difference $\triangle{U} \equiv U-U_N$ reads:
%%%%%
$$
\triangle{U} \approx U_N \frac{15}{8\pi^3}\left(\frac{a}{R}\right)^2 \frac{2\pi}{3} \sum\limits_{k=1}^{+\infty}
\frac{1}{k^2} = U_N \frac{5}{24}\left(\frac{a}{R}\right)^2\, ,
$$
%%%%%
where we took into account $\sum_{k=1}^{\infty} 1/k^2 = \pi^2/6$.
Therefore, this difference is suppressed by power law (with respect to ratio between $a$ and $R$) but not exponentially as it is usually expected
for Kaluza-Klein modes.
Nevertheless, for Sun with $R\approx 7\cdot 10^{10}$cm this value is a negligible part of $U_N$ even if $a\approx 10^{-3}$cm :
$\triangle U \approx 4\cdot 10^{-29}U_N$ where $U_N \approx -2\cdot 10^{48}$erg \cite{Kittel}.

{\em Gravitational interaction of two spheres} \quad Energy of gravitational interaction between two spheres of constant
volume densities $\rho,\rho'$, masses $m,m'$ and radiuses $R,R'$ on a distance $r_3\geq R+R'$ reads
%%%%%%%
\ba{15}
&\phantom{}&U(r_3)=-\frac{G_N mm'}{r_3}\left\{ 1 +18\left(\frac{a}{2\pi R}\right)^3 \left(\frac{a}{2\pi R'}\right)^3 \right.\nn \\
&\times& \left. \sum_{k=1}^{+\infty}\frac{1}{k^6}\exp\left(-\frac{2\pi k}{a}r_3\right)h_k(R)h_k(R')\right\} \, .
\ea
%%%%%%%
%where
%%%%%%
%$$
%h_k(R)=\frac{2\pi|k|R}{a}\cosh\left(\frac{2\pi|k|R}{a}\right)-\sinh\left(\frac{2\pi|k|R}{a}\right)\, .
%$$
%%%%%%
The member of the series with $k=1$ (first Kaluza-Klein mode) gives the main correction to the newtonian expression
%gives the main contribution which
and acquires
the form of Yukawa potential. In this case, for the force of gravitational interaction between these two spheres we obtain:
%%%%%
\ba{16}
-\frac{dU}{dr_3}&\approx&-\frac{G_N mm'}{r_3^2}\left\{ 1 +\frac92\left(\frac{a}{2\pi R}\right)^2 \left(\frac{a}{2\pi R'}\right)^2
\frac{2\pi r_3}{a}\right.\nn \\
&\times& \left. \exp\left[-\frac{2\pi}{a}\left(r_3-R-R'\right)\right]\right\}\, .
\ea
%%%%%
where we made additional natural assumption $R,R'>>a$. If
$r-R-R'\approx a$, then we get an estimate:
%%%%%
\ba{17}
&-&\frac{dU}{dr_3}\approx-\frac{G_N mm'}{r_3^2}\left\{ 1\phantom{\int} \right.\\
&+& \left.0.0084 \left(\frac{a}{2\pi R}\right)^2 \left(\frac{a}{2\pi R'}\right)^2\frac{2\pi r_3}{a}\right\} \nn\, .
\ea
%%%%%
For example, in the case $2\pi R =2\pi R'=10$cm and $a\approx 10^{-2}$cm the correction is $1.68\cdot 10^{-11}$, which is difficult to observe
 at experiments. However, in the case of the internal space topology $T^{d}$
 %where some of periods $a_i$ of the torus are equal each other,
the correction
term in \rf{17} acquires a prefactor $s$ which satisfies the condition $1\leq s\leq d$ and represents a number of extra dimensions with periods of the torus which are equal (or approximately equal) to $a=\max a_i$.
If all of $a_i$ are equal to each other,
%(or approximately equal),
$s=d$. Increasing the number of extra dimensions $d$, finally we arrive at the condition when the correction term becomes big enough to contradict
with experimental data. Therefore, in this case we can get
a limitation  on a maximal number of
extra dimensions for considered models. Certainly, the models with infinite number of extra dimensions with $a=\max a_i$ are forbidden.

{\em Conclusions}\quad We have considered generalization of the Newton's potential to the case of extra dimensions where multidimensional space
has topology $M_D=\mathbb{R}^3\times T^{d}$. It was shown that for model with one extra dimensions in the case of massive point source, the gravitational
potential can be expressed via compact and elegant formula \rf{6}. This formula is valid for full range of variables $r_3$ (magnitude of a radius vector in three
dimensions) and $\xi$ (extra dimension) and has well known asymptotic behavior. Then, this formula was applied to an infinitesimally thin shell, a shell, a
sphere and two spheres to get gravitational potentials and acceleration of a test body for these configurations and to compare obtained results
with the known newtonian formulas. In some cases, obtained potentials and accelerations have logarithmic divergences near the boundaries of shells and spheres. Additionally, in contrast to the newtonian case, test bodies accelerate inside of shells. For each considered problems, we found deviations from the known newtonian expressions and show that for proposed parameters of the models it is difficult to observe these deviations at experiments.
Nevertheless, if internal space has topology $T^d$ with approximately equal periods $a_i$ of the torus in different dimensions, then we can get a limitation on
maximal number of extra dimensions from experimental data with the help of formulas \rf{16},\rf{17} for gravitational interaction between two massive spheres. To conclude this letter, we want to stress that with the help of exact formulas of the form of \rf{6} we can solve multidimensional quantum Schr$\ddot{\mbox{o}}$dinger equation. It opens a possibility to find a new gravitationally bound states: "dark matter atoms". We postpone this investigation to our forthcoming paper.

%%%%%%%%%%%%%%%%%%%%%%%%%%%%%%%%%%%%%%%%%%%%%%%%%%%%%%%%%%%%%%%%%%%%%%%%%%%%%%
%%%%%%%%%%%%%%%%%%%%%%%%%%%%%%%%%%%%%%%%%%%%%%%%%%%%%%%%%%%%%%%%%%%%%%%%%%%%%%5
%\section{\label{sec:7}Summary and discussion}
%\setcounter{equation}{0}

%%%%%%%%%%%%%%%%%%%%%%%%%%%%%%%%%%%%%%%%%%%%%%%%
%\section*{Acknowledgements}
\indent \indent A. Zh. acknowledges the hospitality
of the Theory Division of CERN during preparation of this work.
This work was supported in part by the
"Cosmomicrophysics" programme of the Physics and Astronomy
Division of the National Academy of Sciences of Ukraine.
%%%%%%%%%%%%%%%%%%%%%%%%%%%%%%%%%%%%%%%%%%%%%%%%%

%%%%%%%%%%%%%%%%%%%%%%%%%%%%%%%%%%%%%%%%%%%%%%%%%%%%%%%%%%%%%%%%%%%%%%%%%%%
%%%%%%%%%%%%%%%%%%%%%%%%%%%%%%%%%%%%%%%%%%%%%%%%%%%%%%%%%%%%%%%%%%%%%%%%%%%%

%%%%%%%%%%%%%%%%%%%%%%%%%%%%%%%%%%%%%%%%%%%%%%%%%%%%%%%%%%%%%%%%%%%%%%%%%%%%%%%
%%%%%%%%%%%%%%%%%%%%%%%%%%%%%%%%%%%%%%%%%%%%%%%%%%%%%%%%%%%%%%%%%%%%%%%%%%%%%%%


\begin{thebibliography}{}



\bibitem{ADD}
 N. Arkani-Hamed, S. Dimopoulos and G. Dvali, Phys. Rev. D {\bf 59}, 086004 (1999);
arXiv:hep-ph/9807344.
%%%%%%
\bibitem{CB}
 P. Callin and C.P. Burgess,  Nucl. Phys. B {\bf 752}, 60 (2006);
arXiv:hep-ph/0511216.
%%%%%%
\bibitem{large}
N. Arkani-Hamed, S. Dimopoulos and G. Dvali, Phys. Lett. B {\bf 429}, 263 (1998);
I. Antoniadis, N. Arkani-Hamed, S. Dimopoulos and G. Dvali, Phys. Lett. B {\bf 436}, 257 (1998).
%%%%%%
\bibitem{PBM}
A.P. Prudnikov, Yu.A. Brychkov and O.I. Marichev, {\it Integrals and Series, vol. 1: Elementary Functions},
(Gordon and Breach Science Publishers, New York, 1986).
%%%%%%
\bibitem{inverse-square law}
E.G. Adelberger, B.R. Heckel and A.E. Nelson, Ann. Rev. Nucl. Part. Sci. {\bf 53}, 77 (2003).
%%%%%%
\bibitem{Hoyle}
C. D. Hoyle et al, Phys. Rev. Lett. {\bf 86}, 1418 (2001); arXive:hep-ph/0011014.
%, U. Schmidt, B. R. Heckel, E. G. Adelberger, J. H. Gundlach, D. J. Kapner, H. E. Swanson
%%%%%
\bibitem{supernova}
S. Hannestad and G. Raffelt, Phys. Rev. Lett. {\bf 87}, 051301 (2001); arXiv:hep-ph/0103201.
%%%%%
\bibitem{Kittel}
Ch.Kittel, W. D. Knight and M. A. Ruderman,
{\it Mechanics: Berkeley Physics Course. vol. 1}, (McGraw-Hill, New York, 1965).
%%%%%%



%%%%%%%%%%%%%%%%%%%%%%%%%%%%%%%%%%%%%%%%%%%%%%%%%%%%%%%%%%%%%%%%%%%%%%%%%%%%%%%%%%%%%%%%%%

\end{thebibliography}
\end{document}